\documentclass[aps,prl,reprint,letterpaper,amsmath,amssymb,longbibliography,superscriptaddress]{revtex4-1}
\usepackage{amsmath}
\usepackage{graphicx}

\begin{document}
\title{Resonant spin amplification meets electron spin resonance in $n$-GaAs}
\author{V.~V.~Belykh}
\email[]{belykh@lebedev.ru}
\affiliation{P.~N.~Lebedev Physical Institute of the Russian Academy of Sciences, 119991 Moscow, Russia}
\author{D.~N.~Sob'yanin}
\email[]{sobyanin@lpi.ru}
\affiliation{P.~N.~Lebedev Physical Institute of the Russian Academy of Sciences, 119991 Moscow, Russia}
\author{A.~R.~Korotneva}
\affiliation{P.~N.~Lebedev Physical Institute of the Russian Academy of Sciences, 119991 Moscow, Russia}
\date{\today}
\begin{abstract}
Periodic excitation of electron spin polarization by consecutive laser pulses in phase with Larmor spin precession about a magnetic field results in resonant spin amplification (RSA). We observe a drastic modification of RSA in $n$-doped bulk GaAs under the influence of external oscillating magnetic field.
We find a double-peaked electron spin resonance instead of a single-peaked resonance expected without optical pumping. The frequency splitting increases linearly with amplitude of field oscillations, while the spin deviation increases quadratically. Moreover, we show that the oscillating field can both significantly suppress RSA and induce new conditions for resonance. Using quaternions to describe spin rotations, we develop a theory that simultaneously considers spin precession, decay, and amplification and reproduces the entire set of the experimental data. Using the radio-frequency field allows one to control the conditions of RSA and achieve fine tuning of average spin polarization without modifying the parameters of optical pumping.
\end{abstract}
\maketitle

\section{Introduction}
The electron spin system in a magnetic field $\mathbf{B}$ is characterized by the Larmor frequency $f_\text{L}$ of spin precession, which also determines the Zeeman energy splitting $E_\text{Z} = 2\pi\hbar f_\text{L}$ between the spin eigenstates. Being subject to an additional radiofrequency (rf) or microwave magnetic field oscillating at a frequency $f_\text{rf}$, the spin system can experience a resonance when $f_\text{rf} = f_\text{L}$, referred to as Electron paramagnetic resonance or Electron spin resonance (ESR) \cite{Poole1996}. In this case the rf field is resonantly absorbed while synchronizing precession of individual spins, i.e., driving precession of the macroscopic spin polarization. ESR can also be detected optically, e.g., by measuring the degree of circular polarization of photoluminescence, which significantly increases sensitivity of the method \cite{Cavenett1981}. Usually, ESR affects only the spin polarization given by the difference between the populations of Zeeman-splitted spin sublevels in thermal equilibrium, which is fairly small. The spin polarization affected by ESR can be enhanced by an additional circularly-polarized optical pumping 
\cite{Hermann1971,Cavenett1981,Belykh2020a} thanks to the optical orientation effect \cite{Meier1984}.

In fact, for the special case of a periodic optical pumping another resonance can appear, referred to as Resonant spin amplification (RSA) \cite{Kikkawa1998}. It takes place when the precessing spin created by a laser pulse comes in phase with the spin created by the next pulse, i.e., when $f_\text{L} = n f_\text{o}$, where  $f_\text{o}$ is the repetition rate of optical pulses and $n$ is an integer number. RSA reveals itself as resonant enhancement of spin polarization and is traditionally detected in pump-probe Faraday/Kerr rotation experiments \cite{Yakovlev2017,Yugova2009,Kugler2011,Griesbeck2012,Hernandez2014,Macmahon2019,Schering2019}. 

In this work, we bring together RSA and ESR resonances by applying a rf field to the $n$-doped bulk GaAs sample periodically pumped by laser pulses. To detect the spin polarization, we measure Faraday rotation of a linearly polarized component of the same laser pulses in a single-beam experiment similarly to the method developed in the works \cite{Saeed2018,Kotur2018}. We find that in the RSA regime the behavior of ESR is unusual. In particular, spin polarization is quadratic on the rf field amplitude and shows a drastically different dependence on $f_\text{rf}$ and $f_\text{L}$. The rf field can either drive spin precession out of the RSA resonance or tune it into the RSA resonance. We develop a theory of the combined RSA-ESR resonance using quaternions to describe spin rotations. The theory results in a single analytical equation that fully reproduces the experimental observations.

\section{Experimental details}
The scheme of the experiment is shown in Fig.~\ref{fig:BDep1V}(a). The sample under study is a 350-$\mu$m-thick Si-doped GaAs bulk wafer with electron concentration $1.4 \times 10^{16}$~cm$^{-3}$ near the metal-insulator transition \cite{Belykh2016,Belykh2019,Belykh2020a}. The sample is placed in the variable temperature cryostat. The temperature in all experiments $T = 6$~K.  Using the permanent magnet placed outside the cryostat on a controllable distance, a constant magnetic field $\mathbf{B}$ up to 15~mT is applied along $x$-axis, perpendicular to the direction of light propagation ($z$-axis) and to the sample normal (Voigt geometry). The optical spin pumping and probing is performed by the same laser beam, the initial polarization of which is slightly elliptical. The circular and linear components of the elliptically polarized beam can be considered as the simultaneous pump and probe, respectively, for the electron spin. We use 2-ps-long optical pulses generated by the Ti:Sapphire laser with repetition rate $76$~MHz. To achieve multiple RSA peaks in the limited range of magnetic fields, we reduce the repetition rate of optical pulses 16 times to $f_\text{o} = 4.8$~MHz (repetition period $T_\text{o}=209$~ns) by selecting single pulses with the acousto-optical pulse picker synchronized with the laser. The Faraday rotation of the linear component of laser polarization after transmission through the sample is analyzed using the Wollaston prism, splitting the beam into the two orthogonally polarized beams of approximately equal intensities, which are registered by the balanced photodetector. The laser wavelength is set to 827~nm.

To observe the effect of ESR on the measured Faraday rotation, the rf magnetic field is applied along the sample normal ($z$-axis) using the small ($\lesssim 1$~mm-inner and $\sim 1.5$~mm-outer diameter) coil near the sample surface. The current through the coil is driven by the function generator, which creates sinusoidal voltage with frequency $f_\text{rf}$ up to 150~MHz and amplitude $U_\text{rf}$ up to 10~V. The generator output is modulated at frequency 100~kHz, so that a 5-$\mu$s train of sinusoidal voltage follows 5~$\mu$s of zero voltage. Using the lock-in amplifier, the signal from the balanced detector is registered synchronously at modulation frequency 100~kHz. Thus, we measure the difference between the Faraday rotation for the rf field switched off and on, which is proportional to the corresponding difference in the $z$ component of the spin polarization, $\Delta S_z$.

\section{Experimental results}
\begin{figure}
\includegraphics[width=0.8\columnwidth]{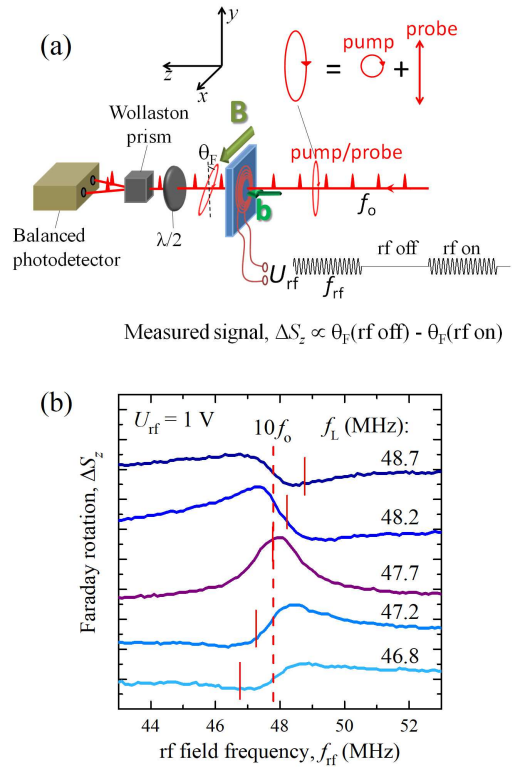}
\caption{(a) Scheme of the experiment. (b) Faraday rotation as a function of the rf field frequency (ESR spectra) for different constant magnetic fields parametrized by the Larmor frequency $f_\text{L}$. The pronounced ESR peak is observed for $f_\text{L} \approx 10 f_\text{o}$. Amplitude of the voltage applied to the rf coil $U_\text{rf} = 1$~V, temperature $T = 6$~K. Red solid lines show the positions of the Larmor frequencies, while red dashed line shows the position of the optical pulses frequency multiplied by 10.}
\label{fig:BDep1V}
\end{figure}

In this study we deal with spins of resident electrons in $n$-doped GaAs. Their spin lifetime at low magnetic fields and low temperatures $\tau_\text{s} = T_2^* \approx T_2 \approx T_1 \sim 200-300$~ns \cite{Belykh2016} is much longer than the recombination time of photocreated electrons and holes ($\lesssim 1$~ns). Here $T_1$, $T_2$, and $T_2^*$ are the longitudinal, transverse, and inhomogeneous transverse spin relaxation times, respectively. When a circularly polarized (component of) laser pulse excites a spin-polarized electron-hole pair, the spin relaxation time of the hole is much shorter than that of the electron and about the recombination time. Thus, the hole with almost equal probability can recombine with the photocreated or the resident electron of any spin polarization, while the spin of the photocreated electron stays in the system. In this way each optical pulse introduces in the system a spin polarization $\Delta S_\text{o}$ directed along the sample normal ($z$-axis) with repetition rate $f_\text{o}$. When the permanent magnetic field $B$ is applied to the sample, the electron spins precess with the Larmor frequency  $f_\text{L} = |g|\mu_\text{B} B /2\pi\hbar$, where $g=-0.44$ is the electron $g$ factor in bulk GaAs and $\mu_\text{B}$ is the Bohr magneton. If the magnetic field fulfills the resonance condition $f_{L} = n f_\text{o}$, where $n$ is an integer, RSA takes place \cite{Kikkawa1998}. RSA for the studied sample was observed in the work \cite{Belykh2016} with the separate pump and probe beams. Here we study a change in the spin polarization $\Delta S_z$ induced by the rf field, in particular in the RSA regime. To do this, we measure $\Delta S_z$ as a function of the rf frequency $f_\text{rf}$, which is referred to as the ESR spectrum.

Figure~\ref{fig:BDep1V}(b) shows the ESR spectra for different values of the permanent magnetic field $B$, corresponding to different $f_\text{L}$ for the relatively small rf voltage $U_\text{rf} = 1$~V.
The Larmor frequency $f_\text{L}$ is changed in the vicinity of the RSA resonance $f_\text{L} = 10 f_\text{o}$. Out of the resonance the ESR spectra exhibit a derivative-like peculiarity at $f_\text{rf} = 10 f_\text{o}$. When $f_\text{L}$ approaches $10 f_\text{o}$, the peculiarity is enhanced and transforms to the pronounced symmetric peak with the half width at half maximum (HWHM) $\delta f_\text{rf} \approx 1$~MHz, which gives an estimate for the spin dephasing time $\tau_\text{s} \sim 1/2\pi\delta f_\text{rf}$ (it will be seen below that, in general, $\delta f_\text{rf}$ is also contributed by the rf field amplitude and $T_\text{o}$). Thus, the effect of rf on spin polarization is most pronounced in the RSA regime.

\begin{figure*}
\includegraphics[width=2\columnwidth]{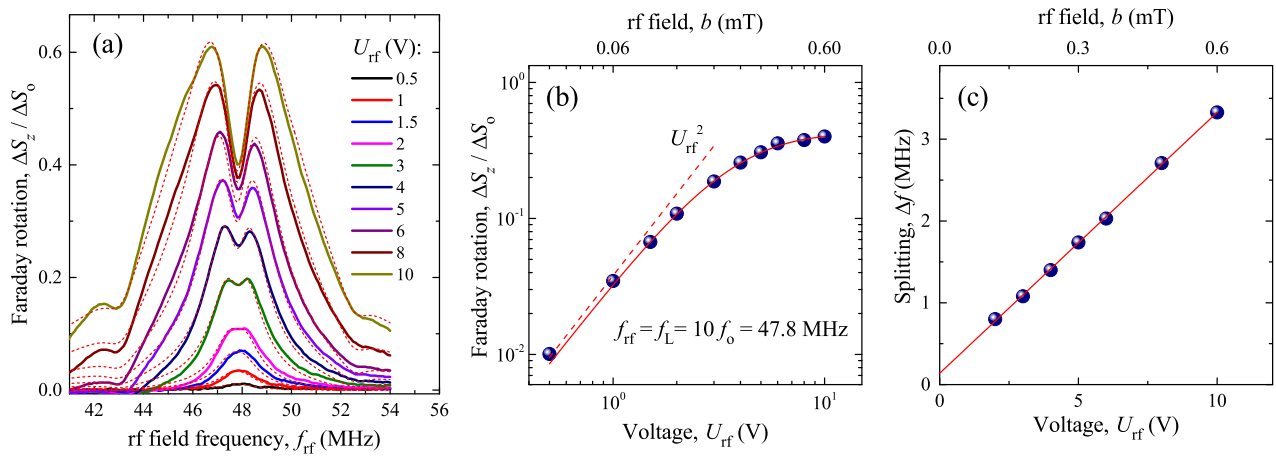}
\caption{Dependence on the rf field amplitude for $f_\text{L} = 10 f_\text{o} = 47.8$~MHz. (a) ESR spectra for different amplitudes of the rf field parametrized by the amplitude of the voltage applied to the rf coil. Red dashed lines show theoretical fits to experimental data. (b) rf voltage dependence of the Faraday rotation signal for $f_\text{rf} = f_\text{L} = 10 f_\text{o}$. Red solid line shows theoretical fit to experimental data, while dashed line shows quadratic dependence. (c) Frequency splitting observed in the ESR spectra as a function of the rf voltage. Red solid line shows linear fit. $T = 6$~K.}
\label{fig:VDep}
\end{figure*}
Next we increase the rf field amplitude $b$ via the voltage applied to the rf coil $U_\text{rf} \propto b$ and observe transformation of the ESR spectrum [Fig.~\ref{fig:VDep}(a)]. These experiments are performed in the RSA regime, $f_\text{L} = 10 f_\text{o}$. With increasing $U_\text{rf}$ the ESR peak amplitude rapidly increases and the peak broadens. At even higher rf voltages the splitting in the ESR peak appears, which we attribute to the Rabi splitting \cite{Rabi1937}. The frequency splitting $\Delta f$ is proportional to $U_\text{rf}$ [Fig.~\ref{fig:VDep}(c)], which is expected for the Rabi splitting. Figure~\ref{fig:VDep}(b) shows the dependence of the spin polarization $\Delta S_z$ on the rf voltage for the resonant case $f_\text{rf} = f_\text{L} = 10 f_\text{o}$. Spin polarization $\Delta S_z$ increases quadratically and then the dependence flattens. Note that, usually the ESR signal linearly increases with $b$ and oscillates at higher $b$ \cite{Belykh2019}.

\begin{figure}
\includegraphics[width=0.7\columnwidth]{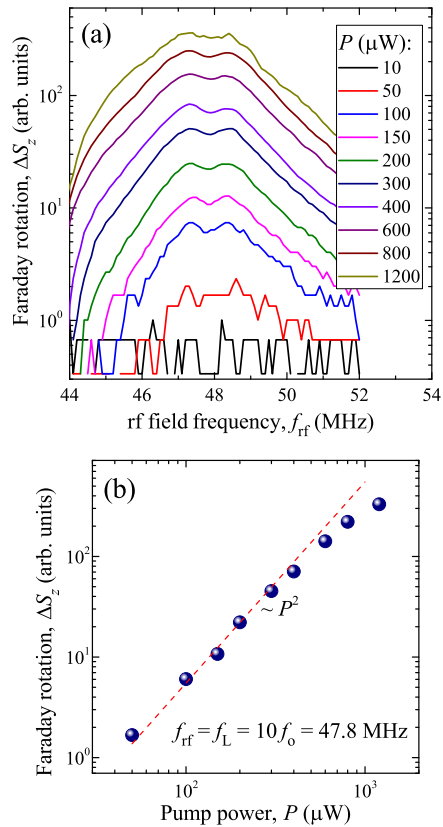}
\caption{Dependence on the optical pump power for $f_\text{L} = 10 f_\text{o} = 47.8$~MHz. (a) ESR spectra for different optical pump powers. Note semilogarithmic scale. (b) Pump power dependence of the Faraday rotation signal for $f_\text{rf} = f_\text{L} = 10 f_\text{o} = 47.8$~MHz. $U_\text{rf} = 4$~V, $T = 6$~K.}
\label{fig:PDep}
\end{figure}
It is instructive to check the behavior of the ESR spectrum with increasing the power of optical pumping $P$ in the RSA regime $f_\text{L} = 10 f_\text{o}$ [Fig.~\ref{fig:PDep}(a)]. The spectrum amplitude strongly increases with $P$, while the form of the spectrum remains almost unchanged. Only at large $P$ the spectrum somewhat broadens, which may be attributed to a decrease of the spin dephasing time at high optical pumping \cite{Crooker2009,Petrov2018}. The signal $\Delta S_z$ at the center of the ESR peak, for $f_\text{rf} = f_\text{L} = 10 f_\text{o}$, increases quadratically with $P$ and slightly saturates at high $P$ [Fig.~\ref{fig:PDep}(b)]. This behavior is expected. Indeed, the spin polarization is proportional, on the one hand, to the pump power and, on the other hand, to the probe power. Here each laser pulse pumps the spin system and simultaneously probes it, so the signal is proportional to $P^2$. And vice versa, the quadratic $P$ dependence confirms that we work in the regime of a one-beam simultaneous pumping and probing, i.e., that optical pumping plays the major role in our experiments.

\begin{figure*}
\includegraphics[width=1.8\columnwidth]{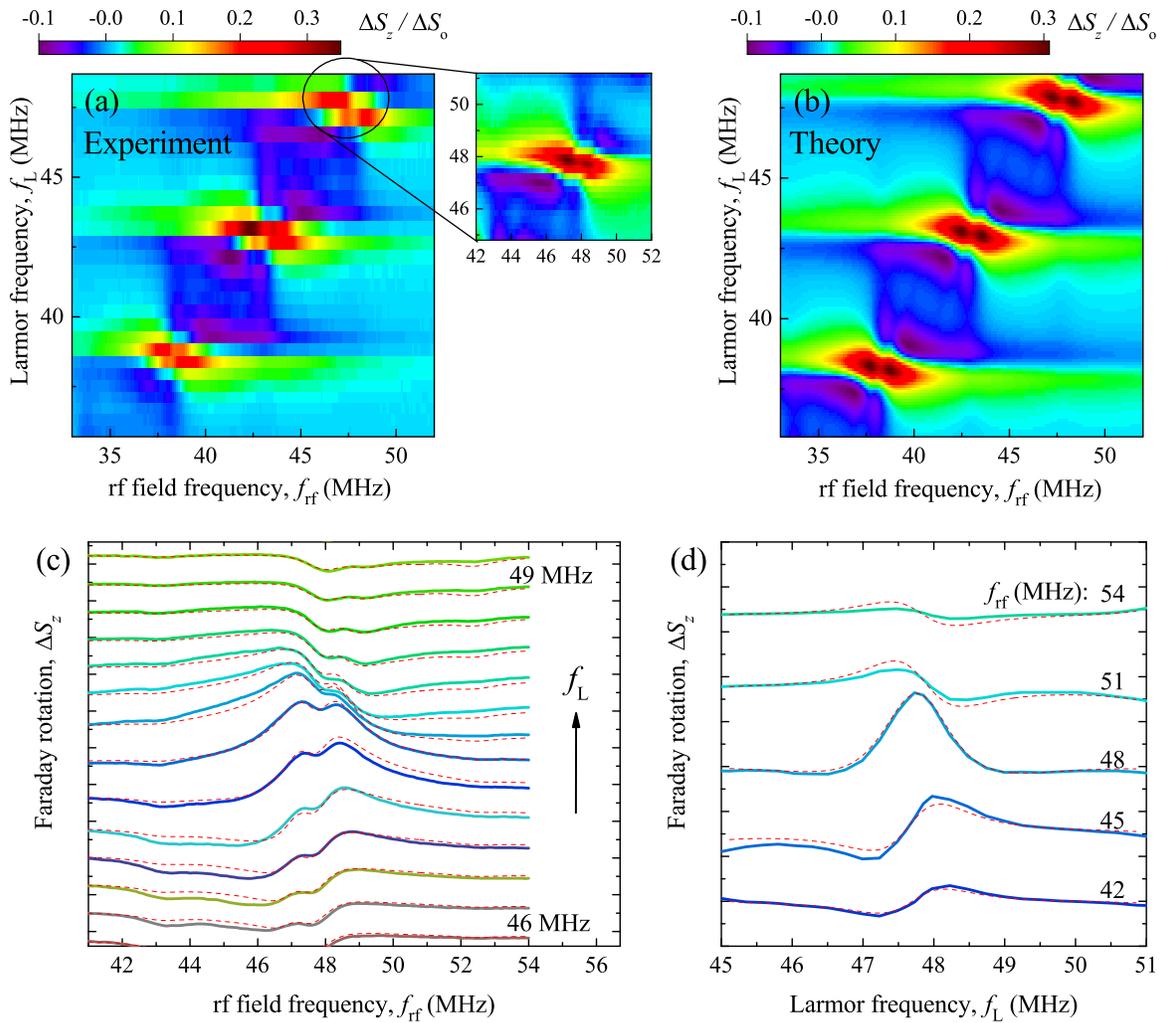}
\caption{(a) Two-dimensional map of the Faraday rotation signal as a function of the rf field frequency and Larmor frequency (magnetic field). Inset shows a closeup of the map. (b) Theoretical calculation of the corresponding map. (c) ESR spectra for different Larmor frequencies. (d) Faraday rotation as a function of the Larmor frequency for different rf field frequencies. In panels (c) and (d) the curves are vertically shifted for clarity and red dashed lines show theoretical fits to experimental data. $U_\text{rf} = 4$~V, $T = 6$~K.}
\label{fig:BDep4V}
\end{figure*}
Finally, we show the comprehensive picture of the spin polarization dependence on $f_\text{rf}$ and $f_\text{L}$ for $U_\text{rf}=4$~V [Fig.~\ref{fig:BDep4V}(a)]. $\Delta S_z$ has pronounced maxima at the ESR-RSA resonances, where $f_\text{rf} = f_\text{L} = n f_\text{o}$. We observe resonances for $n=8$, $9$, and $10$, remind that $f_\text{o} = 4.8$~MHz. Interestingly, we do not observe any resonance features on the line $f_\text{rf} = f_\text{L}$ outside the resonances with $f_\text{o}$. This indicates that the pure ESR on the nonamplified spin polarization is relatively small, while RSA dramatically enhances the ESR effect. Later on we will show that this observation can be treated the other way round as suppression of RSA by the resonant rf field. Another observation that can be made is that the width of the resonance is larger than its height, i.e., $\delta f_\text{rf} > \delta f_\text{L}$. This is also seen in Figs.~\ref{fig:BDep4V}(c) and~\ref{fig:BDep4V}(d) showing the dependences of spin polarization on $f_\text{rf}$ and $f_\text{L}$, respectively (horizontal and vertical slices of the map in Fig.~\ref{fig:BDep4V}(a) near the $n=10$ resonance). The width $\delta f_\text{rf}$ of $\Delta S_z (f_\text{rf})$ for $f_\text{L} = n f_\text{o}$ is determined by the spin dephasing time $\tau_\text{s}$ and, as follows from Fig.~\ref{fig:VDep}(a), by the rf field amplitude. On the other hand, the corresponding width $\delta f_\text{L}$ seems to be contributed by $\tau_\text{s}$ and independent of the rf field amplitude. In fact, $T_\text{o}$ also contributes to the spectral widths for $T_\text{o} \sim \tau_\text{s}$, as it follows from the theory below. The spectra in Fig.~\ref{fig:BDep4V}(c) and Fig.~\ref{fig:BDep4V}(d) show similar derivative-like peculiarity at frequency $10 f_\text{o}$ when $f_\text{L}$ or $f_\text{rf}$, respectively, is out of resonance with $10 f_\text{o}$.

\section{Theory}

The behavior of the total electron spin $\mathbf{S}$ in an external magnetic field under RSA is described by an inhomogeneous Bloch equation
\begin{equation}
\label{Bloch}
\frac{d\mathbf{S}}{dt}=\boldsymbol{\omega}\times\mathbf{S}-\frac{\mathbf{S}}{\tau_\text{s}}+\Delta\mathbf{S}_\text{o}\sum_n\delta(t-nT_\text{o}),
\end{equation}
where $\tau_\text{s}$ is the spin relaxation time at low $B$ and $\boldsymbol{\omega}=\boldsymbol{\omega}_\text{L}+\mathbf{\Omega}_\text{R}$ is the angular velocity of spin precession, which is the sum of a constant component $\boldsymbol{\omega}_\text{L}=\omega_\text{L}\mathbf{e}_x$ parallel to the permanent magnetic field $\mathbf{B}$, with $\omega_\text{L}=g \mu_\text{B} B /\hbar$ being the Larmor frequency, and a rotating component $\mathbf{\Omega}_\text{R}=\Omega_\text{R}(\mathbf{e}_y\cos\phi+\mathbf{e}_z\sin\phi)$ orthogonal to $\mathbf{B}$, with $\Omega_\text{R}= g \mu_\text{B} b/2\hbar$ being the Rabi frequency, $\phi=\phi_0+\omega_\text{rf}t$, and $\omega_\text{rf}=2\pi f_\text{rf}\operatorname{sgn}g$. Recall that the experimental oscillating rf magnetic field is the sum of two fields rotating in opposite directions with frequency $\omega_\text{rf}$, and the counter-rotating term, i.e., rotating oppositely to $\boldsymbol{\omega}_\text{L}$, is neglected as $b\ll B$ \cite{Abragam1961}. The inhomogeneous term in Eq.~\eqref{Bloch} with Dirac delta functions $\delta(t-nT_\text{o})$ describes repetitive amplification of $z$ spin component by successive equidistant laser pulses arriving at times $nT_\text{o}$  by a quantity $\Delta S_\text{o}$ per pulse, so that $\Delta\mathbf{S}_\text{o}=\mathbf{e}_z\Delta S_\text{o}$. The term with thermal equilibrium spin is dropped off from Eq.~\eqref{Bloch} because it is negligible compared to the resulting spin polarization achieved in the experiment under optical pumping.

Between every two adjacent pulses the spin evolves according to the homogeneous Bloch equation, Eq.~\eqref{Bloch} without the last term. Multiplying this equation scalarly by $\mathbf{S}$ removes the term $\boldsymbol{\omega}\times\mathbf{S}$ and reveals an exponential decay of the absolute spin magnitude, $S=s e^{-t/\tau_\text{s}}$, so it is natural to represent the spin vector in the form of the product
\begin{equation}
\label{spinDecay}
\mathbf{S}=\mathbf{s}e^{-t/\tau_\text{s}},
\end{equation}
where the vector amplitude $\mathbf{s}$ satisfies
\begin{equation}
\label{vectorSpinAmplitude}
\frac{d\mathbf{s}}{dt}=\boldsymbol{\omega}\times\mathbf{s},
\end{equation}
as one may verify with substituting Eq.~\eqref{spinDecay} in the homogeneous Bloch equation. Equation \eqref{vectorSpinAmplitude} means that only the orientation of $\mathbf{s}$, not its absolute value $s$, changes with time. Thus, by Eq.~\eqref{spinDecay} the spin rotates and simultaneously decays until the next laser pulse arrives, and the rotation is fully determined by the angular velocity~$\boldsymbol{\omega}$.

The current spin orientation is characterized by the unit vector $\mathbf{e}=\mathbf{e}(t)=\mathbf{s}/s$ and can be obtained from the initial spin orientation $\mathbf{e}_0=\mathbf{e}(0)$ by rotation around a unit axis $\boldsymbol{\zeta}=\boldsymbol{\zeta}(t)$ through an angle $\alpha=\alpha(t)$, which can be expressed thus:
\begin{equation}
\label{spinRotation}
\mathbf{e}=\Lambda\circ\mathbf{e}_0\circ\bar{\Lambda},
\end{equation}
where
\begin{equation}
\label{quaternionLambda}
\Lambda=\cos\frac{\alpha}{2}+\boldsymbol{\zeta}\sin\frac{\alpha}{2}=e^{\boldsymbol{\zeta}\alpha/2}
\end{equation}
is a quaternion, a special type of hypercomplex numbers particularly effective in various physical problems involving rotations \cite{Sobyanin2016,GorhamLaughlin2019,KadekEtal2019,GubbiottiEtal2019,AkemannEtal2019,FaureBeaulieuNoiray2020}. A quaternion $\mathrm{M}=\mu_0+\boldsymbol{\mu}$ is the sum of a scalar part $\mu_0$ and a vector part $\boldsymbol{\mu}$, and the corresponding conjugate quaternion is $\bar{\mathrm{M}}=\mu_0-\boldsymbol{\mu}$. Any vector is a quaternion with zero scalar part; e.g., the spin $\mathbf{S}$ can be considered as the quaternion $0+\mathbf{S}$. The product of two quaternions $\mathrm{M}$ and $\mathrm{N}=\nu_0+\boldsymbol{\nu}$ is defined as
\begin{equation}
\label{quaternionProduct}
\mathrm{M}\circ \mathrm{N}=\mu_0\nu_0-\boldsymbol{\mu}\cdot\boldsymbol{\nu}+\mu_0\boldsymbol{\nu}+\nu_0\boldsymbol{\mu}+\boldsymbol{\mu}\times\boldsymbol{\nu}
\end{equation}
and is associative, $(\mathrm{\Lambda}\circ\mathrm{M})\circ\mathrm{N}=\mathrm{\Lambda}\circ(\mathrm{M}\circ\mathrm{N})=\mathrm{\Lambda}\circ\mathrm{M}\circ\mathrm{N}$, but not commutative, $\mathrm{M}\circ\mathrm{N}\neq\mathrm{N}\circ\mathrm{M}$. Any quaternion of the form \eqref{quaternionLambda} has unit norm, which means that $\Lambda\circ\bar{\Lambda}=\bar{\Lambda}\circ\Lambda=1$. It follows from Eq.~\eqref{quaternionLambda} that for any quaternion $\mathrm{M}$ of unit norm
\begin{equation}
\label{wrapping}
\exp\Bigl(\mathrm{M}\circ\frac{\boldsymbol{\zeta}\alpha}2\circ\bar{\mathrm{M}}\Bigr)=\mathrm{M}\circ e^{\boldsymbol{\zeta}\alpha/2}\circ\bar{\mathrm{M}}.
\end{equation}

The knowledge of the time behavior of $\Lambda=\Lambda(t)$ means the knowledge of the rotational dynamics of the spin. In analogy to the work \cite{BelykhEtal2020}, one can write
\begin{equation}
\label{Lambda}
\Lambda=\mathrm{M}\circ\mathrm{N},
\end{equation}
where $\mathrm{M}=e^{\boldsymbol{\omega}_\textrm{rf}t/2}$ and $\mathrm{N}=e^{\mathbf{\Omega}_0t/2}$, with $\boldsymbol{\omega}_\text{rf}=\omega_\text{rf}\mathbf{e}_x$. The vector $\mathbf{\Omega}_0=\mathbf{\Omega}(0)=-\Delta\omega_\text{rf}\mathbf{e}_x+\Omega_\text{R}\cos\phi_0\mathbf{e}_y+\Omega_\text{R}\sin\phi_0\mathbf{e}_z$, where $\mathbf{\Omega}=\boldsymbol{\omega}-\boldsymbol{\omega}_\text{rf}$, has absolute value
$\Omega=\sqrt{\Delta\omega_\text{rf}^2+\Omega_\text{R}^2}$, where $\Delta\omega_\text{rf}=\omega_\text{rf}-\omega_\text{L}$ is the angular frequency detuning. Equation~\eqref{Lambda} shows that the current spin orientation is obtained from the initial orientation with superposition of two successive rotations around fixed axes, the first around $\mathbf{\Omega}_0/\Omega$ through angle $\Omega t$ and the second around $\mathbf{e}_x$ through angle $\omega_\text{rf}t$. Both rotations are performed in the laboratory frame of reference, but an interpretation invoking the rotating frame of reference is evident: the first rotation corresponds to spin precession with angular velocity $\mathbf{\Omega}_0$ in the rotating frame, while the second rotation describes proper rotation with angular velocity~$\boldsymbol{\omega}_\text{rf}$ of the frame itself, thus giving the spin orientation already in the laboratory frame.

Now we may return to the full dynamics and calculate the observed spin signal. The pump-probe spectroscopy measures the spin (more precisely, its $z$ component) and simultaneously amplifies it, but this occurs only at the times of arrival of laser pulses, when the spin changes by $\Delta\mathbf{S}_\text{o}$. Assuming that the spin measurement occurs just before its amplification, we may relate the spins at two adjacent times of measurement,
\begin{equation}
\label{adjacentSpins}
\mathbf{S}_{n+1}=\Lambda_n\circ(\mathbf{S}_n+\Delta\mathbf{S}_\text{o})\circ\bar{\Lambda}_n e^{-T_\text{o}/\tau_\text{s}},
\end{equation}
where $\Lambda_n$ is the quaternion corresponding to spin rotation between the $n$th and $(n+1)$th pulses. Equation \eqref{adjacentSpins} equally relates $\mathbf{S}_n$ and $\mathbf{S}_{n-1}$, $\mathbf{S}_{n-1}$ and $\mathbf{S}_{n-2}$, and so on, so repetitively iterating gives the spin at some time of measurement (we arbitrarily choose $t=0$),
\begin{equation}
\label{iterativeSpin}
\mathbf{S}_0=\sum_{n=1}^\infty\mathrm{Q}_n\circ\Delta\mathbf{S}_\text{o}\circ\bar{\mathrm{Q}}_n e^{-nT_\text{o}/\tau_\text{s}},
\end{equation}
where
\begin{equation}
\label{Qn}
\mathrm{Q}_n=\Lambda_{-1}\circ\ldots\circ\Lambda_{-n}
\end{equation}
is the quaternion giving rotation starting at the time $n$ pulses before and ending at the time of measurement.

Every $\Lambda_{-i}$ entering Eq.~\eqref{Qn} corresponds to spin rotation from the $-i$th to the $(-i+1)$th pulse and is given by Eq.~\eqref{Lambda} in which we should put $t=T_\text{o}$ and take as $\mathbf{\Omega}_0$ the initial orientation of $\mathbf{\Omega}$ at time $-iT_\text{o}$, which is denoted by $\mathbf{\Omega}_{-i}=\mathbf{\Omega}(-iT_\text{o})$. The vector $\mathbf{\Omega}$ rotates around $\mathbf{e}_x$ with frequency $\omega_\text{rf}$, which by Eq.~\eqref{spinRotation} is expressed as $\mathbf{\Omega}=\mathrm{M}\circ\mathbf{\Omega}_0\circ\bar{\mathrm{M}}$. Therefore, $\mathbf{\Omega}_{-i}=\bar{\mathrm{M}}_0^i\circ\mathbf{\Omega}_0\circ\mathrm{M}_0^i$, where $\mathrm{M}_0=e^{\boldsymbol{\omega}_\textrm{rf}T_\text{o}/2}$ and $\mathrm{M}^n$ denotes the $n$th power of a quaternion $\mathrm{M}$, the quaternionic product of $n$ identical quaternions, $\mathrm{M}^n=\mathrm{M}\circ\ldots\circ\mathrm{M}$. We then have $\Lambda_{-i}=\mathrm{M}_0\circ\mathrm{N}_{-i}$, where by Eq.~\eqref{wrapping} $\mathrm{N}_{-i}=e^{\mathbf{\Omega}_{-i}T_\text{o}/2}=\bar{\mathrm{M}}_0^i\circ\mathrm{N}_0\circ\mathrm{M}_0^i$ and $\mathrm{N}_0=e^{\mathbf{\Omega}_0T_\text{o}/2}$, so that
\begin{equation}
\label{LambdaMinusI}
\Lambda_{-i}=\bar{\mathrm{M}}_0^{i-1}\circ\mathrm{N}_0\circ\mathrm{M}_0^i.
\end{equation}
Combining Eqs.~\eqref{Qn} and \eqref{LambdaMinusI} yields $\mathrm{Q}_n=\mathrm{N}_0\circ\mathrm{M}_0\circ\bar{\mathrm{M}}_0\circ\mathrm{N}_0\circ\mathrm{M}_0^2\circ\bar{\mathrm{M}}_0^2\circ\mathrm{N}_0\circ\mathrm{M}_0^3\circ\ldots\circ\bar{\mathrm{M}}_0^{n-1}\circ\mathrm{N}_0\circ\mathrm{M}_0^n$, or finally
\begin{equation}
\label{QnFinal}
\mathrm{Q}_n=\mathrm{N}_0^n\circ\mathrm{M}_0^n.
\end{equation}
We see from Eqs.~\eqref{spinRotation} and \eqref{QnFinal} that $\mathrm{Q}_n$ describes rotation of $\Delta\mathbf{S}_\text{o}$, initially oriented along~$\mathbf{e}_z$, around $\mathbf{e}_x$ through angle $n\omega_\text{rf}T_\text{o}$ and subsequent rotation around $\mathbf{\Omega}_0/\Omega$ through angle $n\Omega T_\text{o}$.

To find the observed $S_z$, we should project Eq.~\eqref{iterativeSpin} on $\mathbf{e}_z$, which can be performed by using Eq.~\eqref{quaternionProduct} or applying the rotation tensor \cite{Wittenburg2016},
\begin{widetext}
\begin{eqnarray}
\label{spinProjection}
S_z=&&\Delta S_\text{o}\sum_{n=1}^\infty\Bigl\{\Bigl[\cos (n\omega_\text{rf}T_\text{o})\cos (n\Omega T_\text{o})+\frac{\Delta\omega_\text{rf}}{\Omega}\sin (n\omega_\text{rf}T_\text{o})\sin (n\Omega T_\text{o})\Bigr]\nonumber\\
&&-\frac{\Omega_\text{R}^2}{\Omega^2}\bigl[1-\cos (n\Omega T_\text{o})\bigr]\sin\phi_0\sin(n\omega_\text{rf}T_\text{o}-\phi_0)\Bigr\}e^{-nT_\text{o}/\tau_\text{s}}.
\end{eqnarray}
In the experiment, we measure the signal averaged over $\phi_0$, more precisely, the difference $\langle\Delta S_z\rangle=\langle S_z\rangle_{b=0}-\langle S_z\rangle$ between the average signal in the absence, $\langle S_z\rangle_{b=0}$, and the presence, $\langle S_z\rangle$, of the rf magnetic field. By transforming the products of trigonometric functions in Eq.~\eqref{spinProjection} to their sums, averaging over $\phi_0$, directly summing the resulting series by using the formula $\sum_n e^{-na}\cos nx=\bigl[\sinh a/(\cosh a -\cos x)-1\bigr]/2$ \cite{GradshteynRyzhik2007}, and calculating and subtracting $\langle S_z\rangle_{b=0}$, we arrive at the final expression for the observed ESR spectrum under RSA,
\begin{eqnarray}
\label{eq:finalSpinSignal}
\langle\Delta S_z\rangle=&&\frac{\Delta S_\text{o}}8\sinh\frac{T_\text{o}}{\tau_\text{s}}\biggl(\frac4{\cosh (T_\text{o}/\tau_\text{s})-\cos(\omega_\text{L}T_\text{o})}-\frac{2\Omega_\text{R}^2/\Omega^2}{\cosh (T_\text{o}/\tau_\text{s})-\cos(\omega_\text{rf}T_\text{o})}\nonumber\\
&&-\frac{\bigl(1-\Delta\omega_\text{rf}/\Omega\bigr)^2}{\cosh (T_\text{o}/\tau_\text{s})-\cos[(\omega_\text{rf}+\Omega)T_\text{o}]}-\frac{\bigl(1+\Delta\omega_\text{rf}/\Omega\bigr)^2}{\cosh (T_\text{o}/\tau_\text{s})-\cos[(\omega_\text{rf}-\Omega)T_\text{o}]}\biggr).
\end{eqnarray}
\end{widetext}

\section{Discussion}
\begin{figure}
\includegraphics[width=0.8\columnwidth]{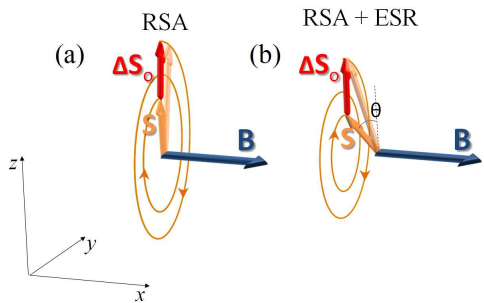}
\caption{(a) Scheme of RSA. Spin polarization precesses about $\mathbf{B}$ and decays. After an integer number of turns it is enhanced by $\Delta S_\text{o}$ via optical pumping.  (b) Effect of ESR on RSA. An oscillating rf field leads to declination of the spin from the $yz$ plane characterized by an increase in angle $\theta$.}
\label{fig:RSAESR}
\end{figure}

Equation \eqref{eq:finalSpinSignal} provides good fit to the experimental data as shown in Figs.~\ref{fig:VDep}(a),~\ref{fig:VDep}(b),~\ref{fig:BDep4V}(c),~\ref{fig:BDep4V}(d). The theory also finely reproduces the two-dimensional map $\Delta S_z (f_\text{rf}, f_\text{L})$ [Figs.~\ref{fig:BDep4V}(a),~\ref{fig:BDep4V}(b)]. In the calculations we use $\tau_\text{s} = 200$~ns. By comparing the experiment to the theory, we can relate the Faraday rotation signal measured in the experiment, $F$, to the rf-induced change in the spin polarization, $\Delta S_z$, measured in units of the spin polarization created by the laser pulse, $\Delta S_\text{o}$, i.e., we can determine the coefficient $k$ in the relation $\Delta S_z / \Delta S_\text{o} = k F$. Similarly, we can determine the coefficient between the rf field amplitude $b$ and rf voltage $U_\text{b}$.  We use the same coefficients for all the data. In this way the signal in Figs.~\ref{fig:VDep}(a),~\ref{fig:VDep}(b),~\ref{fig:BDep4V}(a) is shown not in arbitrary units but directly in units of $\Delta S_\text{o}$, while the upper scale in Figs.~\ref{fig:VDep}(b) and \ref{fig:VDep}(c) shows the values of $b$.

Thus, we observe clear effect of ESR on RSA, and all experimental dependences are well reproduced by the single analytical equation \eqref{eq:finalSpinSignal} with the same parameters. To understand the effect qualitatively, we refer to the scheme in Fig.~\ref{fig:RSAESR}. In the absence of the rf field [Fig.~\ref{fig:RSAESR}(a)], spin $\mathbf{S}$ precesses in the $yz$ plane about magnetic field $\mathbf{B}$ with frequency $f_\text{L}$, decays with time $\tau_\text{s}$ and is enhanced by $\Delta S_\text{o}$ with period $T_\text{o}$. For RSA $f_\text{L} = n f_\text{o}$ and we can write $S_\text{RSA} = (S_\text{RSA} + \Delta S_\text{o}) \exp(-T_\text{o}/\tau_\text{s})$, so
\begin{equation}
S_\text{RSA} = \Delta S_\text{o} / [\exp(T_\text{o}/\tau_\text{s})-1] \approx 0.6 \Delta S_\text{o}.
\label{eq:SRSA}
\end{equation}
When the oscillating rf field with frequency $f_\text{rf} = f_\text{L}$ is applied to the system, it leads to declination of the spin from the $yz$ plane by an angle $\theta \sim \Omega_\text{R} T_\text{o}$, while the optical pumping drives the spin back toward the $yz$ plane [Fig.~\ref{fig:RSAESR}(b)]. As a result, the measured $z$ component acquires a factor $\cos(\theta)$. In the experiment we measure the reduction of the spin polarization $\Delta S_z \sim S_\text{RSA} (1 - \cos\theta) \sim S_\text{RSA} \Omega_\text{R}^2 T_\text{o}^2 / 2 \propto b^2$ for small $b$. These rough estimates explain the unexpected quadratic increase in $\Delta S_z$ with $U_\text{rf}$ observed in Fig.~\ref{fig:VDep}(b). This is contrary to the classical ESR case, where $\mathbf{S}$ is aligned along $\mathbf{B}$ while the rf field drives it out this direction, and the precessing spin component that is measured in the experiment is proportional to $\sin(\theta) \propto b$ for small $b$.

In this way, the resonant rf field inhibits RSA. Figure~\ref{fig:VDep}(a) shows that at the maximal rf voltage corresponding to $b = 0.6$~mT [upper scale in Fig.~\ref{fig:VDep}(b)] the maximal inhibition of the spin signal is achieved at $f_\text{rf} \approx f_\text{L} \pm \Omega_\text{R} / 2\pi$ and reaches $\Delta S_z \approx 0.6 \Delta S_\text{o}$. This means that the rf field almost completely suppresses $S_\text{RSA}$ [see Eq.~\eqref{eq:SRSA}]. On the other hand, in the color map [Fig.~\ref{fig:BDep4V}(a) and (b)] one can note the negative signal $\Delta S_z$, which means the enhancement of RSA. The minimal $\Delta S_z$ is reached slightly out of the RSA resonance $f_\text{L} = n f_\text{o}$ and can be interpreted as the tuning of spin precession toward the RSA condition with the rf field. Thus, the rf field not only inhibits spin polarization in the RSA condition but also recovers spin polarization out of RSA. As follows from Fig.~\ref{fig:BDep4V}(a), at $U_\text{rf} = 4$~V the maximal spin inhibition is about $0.3 \Delta S_\text{o}$, while the maximal spin recovery is about $0.1 \Delta S_\text{o}$.

\section{Conclusions}

We have studied the combined RSA-ESR resonance by pumping and simultaneously probing the electron spin polarization in $n$-GaAs by periodic laser pulses and applying an oscillating rf magnetic field. We have shown that the rf field strongly modifies the optically-created and amplified spin polarization. At small rf field amplitudes, this modification is maximal at the ESR condition and quadratic on the rf field amplitude. At higher rf field amplitudes, the spin polarization shows the double-peaked behavior as a function of the rf field frequency. These observations are unusual when compared to the standard ESR. The rf field can both drive spin precession out of the RSA resonance, thus suppressing spin polarization, or enhance it by tuning spin precession towards the RSA resonance. The experimental results are fully reproduced by a theory considering the electron spin precession in a varied magnetic field under pulsed illumination. This study paves the way for the control and manipulation of the optical spin amplification by the rf or microwave field.

\section{Acknowledgments}
\begin{acknowledgments}
We are grateful to I.~A.~Akimov for fruitful discussions. The work was supported by the Russian Science Foundation through grant No. 18-72-10073.
\end{acknowledgments}

\end{document}